\documentstyle[12pt]{article}
\newcommand{\beq}{\begin{equation}}
\newcommand{\eeq}{\end{equation}}
\newcommand{\beqa}{\begin{eqnarray}}
\newcommand{\eeqa}{\end{eqnarray}}
\newcommand{\beqar}{\begin{eqnarray*}}
\newcommand{\eeqar}{\end{eqnarray*}}
\newcommand{\al}{\alpha}
\newcommand{\lam}{\lambda}
\newcommand{\veps}{\varepsilon}

\newcommand{\G}{\Gamma}

\renewcommand{\l}{\lambda}

\newcommand{\s}{\sigma}

\newcommand{\cH}{{\cal H}}

\newcommand{\eg}{{\it e.g.,}\ }
\newcommand{\ie}{{\it i.e.,}\ }

\newcommand{\norm}[1]{\raise.3ex\hbox{:}#1\raise.3ex\hbox{:}}

\newcommand{\labell}[1]{\label{#1}} 
\newcommand{\reef}[1]{(\ref{#1})}
\newcommand{\Tr}{{\rm Tr}}
\newcommand{\STr}{{\rm STr}}

\newcommand\hi{{\rm i}}

\newcommand\prt{\partial}
\newcommand\wdg{\wedge}
\newcommand\ls{\ell_s}

\newcommand\tR{{\widetilde R}}

\newcommand\Cf{C^{(4)}}

\newcommand\hR{{\hat{R}}}
\newcommand\signot{\sigma_\infty}
\newcommand\E{{\cal E}}
\begin{document}

\thispagestyle{empty}
\rightline{\small hep-th/9911136 \hfill McGill/99-34}
\rightline{\small\hfill NSF-ITP-99-138}
\vspace*{2cm}

\begin{center}
{\bf \LARGE The Noncommutative Bion Core}
\vspace*{1cm}

Neil R. Constable\footnote{E-mail: constabl@hep.physics.mcgill.ca},
Robert C. Myers\footnote{E-mail: rcm@hep.physics.mcgill.ca}
and \O yvind Tafjord\footnote{E-mail: tafjord@physics.mcgill.ca}

\vspace*{0.2cm}

{\it Institute for Theoretical Physics, University of California}\\
{\it Santa Barbara, CA 93106, USA}\\[.5em]

{\it Department of Physics, McGill University}\\
{\it Montr\'eal, QC, H3A 2T8, Canada}\footnote{Permanent address}\\
\vspace{2cm} ABSTRACT
\end{center}

We examine noncommutative solutions of the nonabelian theory
on the world-volume of $N$ coincident D-strings. These solutions
can be interpreted in terms of noncommutative geometry as funnels
describing the nonabelian D-string expanding out into an orthogonal D3-brane.
These configurations are `dual' to the bion solutions in the
abelian world-volume theory of the D3-brane. In the latter, a
charge $N$ magnetic monopole describes $N$ D-strings attached
to the D3-brane with a spike deformation of the world-volume.
The noncommutative D-string solutions give a reliable account of
physics at 
the core of the monopole, where the bion description is expected
to breakdown. In the large $N$ limit,
we find good agreement between the two points of view, including
the energy, couplings to background fields, and the shape of the
funnel.
We also study fluctuations traveling along the D-string, again
obtaining agreement in the large $N$ limit. At finite $N$, our results
give a limit on the number of modes that can
travel to infinity along the $N$ D-strings attached to the D3-brane.

\vfill \setcounter{page}{0} \setcounter{footnote}{0}
\newpage

\section{Introduction}

D-branes \cite{Polchin, leigh, Polchin2} 
have become important tools in the quest to
develop a full understanding of string theory. The low
energy action describing the dynamics of test D-branes consists
of two parts: the Born-Infeld action \cite{bin} and the
Chern-Simons action \cite{mike,cs}. This nonlinear action 
reliably captures the physics of D-branes with great accuracy. 
One interesting aspect of this story is that
one finds that the D-brane action supports solitonic
configurations describing
lower-dimensional branes protruding from the original D-brane 
\cite{calmald, gibb, selfd}.

For instance, in the case of a D3-brane,
one finds spike solutions corresponding to
fundamental strings and D-strings (as well as 
strings) attached to the D3-brane. These configurations have both the 
world-volume gauge fields
and transverse scalar fields excited. The gauge field corresponds to
that of a point charge arising from the end-point of the attached
string, {\it i.e.}, an electric charge for a fundamental
string and a magnetic monopole charge
for a  D-string. The scalar field
on the other hand, represents a deformation of the geometry of the D3-brane,
caused by attaching the strings.

Naively, the range of validity of this analysis is limited to a range
far from the core of the spike where the fields on the D3-brane
world-volume are slowly varying. This range can be increased by 
increasing the number $N$ of attached
strings. (Although $N$ can not be too large if we are to ignore
gravitational effects). However, the results obtained seem to have a
larger regime of validity, maybe even all the way to center of the
spike where it protrudes an infinite distance from the
original position of the D3-brane. This
can partly be understood in light of the fact that the basic BPS spike
is a solution to the full derivative-corrected equations of motion
following from string theory \cite{larus}. Even the dynamics of the
spike, as probed through small fluctuations, agree with expected
string behavior \cite{larspeet,fluct,K/T}. However,
Kastor and Traschen \cite{K/T} showed that certain fluctuation modes
that are inherently three-dimensional also appear propagate to infinity in
this picture, and hence the spike seems to retain
its three-brane character even at large distances.

The purpose of this paper is to study the `dual' description of a
system of $N$ D-strings attached to a D3-brane. This system has been
analyzed previously in
\cite{ded,gauntlett,giveon,sethi,hashimoto,gorsky} using the
connection \cite{ded} 
between the Nahm equations for BPS monopoles \cite{nahm}
and the BPS condition for the nonabelian D-string theory.
This theory contains
noncommutative solutions describing the D-strings expanding out in a
funnel-like geometry to become an orthogonal D3-brane. These
solutions are valid in a regime complimentary to the bion spikes
discussed above. That is, the solutions will accurately describe
the physics very close to the center of the spike, or alternatively
very far from the D3-brane. The two approaches, \ie the D3-brane
spikes and the D-string funnels, turn out to agree exactly in the
large $N$ limit, while we get new insights into the physics at finite
$N$ near the core of the spike from the D-string funnels.

In the next section we will quickly review the full nonabelian
D-brane action, followed by an outline of the remainder of this paper.

\vfill
\newpage
\section{Nonabelian Brane Action}

Our starting point is the nonabelian world-volume action describing $N$
coincident D-strings, whose complete form was recently discussed by Myers
\cite{dielec}, as well as Taylor and Van Raamsdonk \cite{watiprep}. The action
consists of two parts: the Born-Infeld action
\beq
{S}_{BI}=-T_1 \int d^{2}\sigma\,\STr\left(e^{-\phi}\sqrt{-\det\left(
P\left[E_{ab}+E_{ai}(Q^{-1}-\delta)^{ij}E_{jb}\right]+
\l\,F_{ab}\right)\,\det(Q^i{}_j)}
\right),
\labell{finalbi}
\eeq
with 
\beq
\lambda=2\pi \ls^2,\qquad 
E_{\mu\nu}=G_{\mu\nu}+B_{\mu\nu},
\qquad{\rm and}\qquad
Q^i{}_j\equiv\delta^i{}_j+i\lambda\,[\Phi^i,\Phi^k]\,E_{kj},
\labell{extra6}
\eeq
and the Chern-Simons action
\beq
S_{CS}=\mu_1\int \STr\left(P\left[e^{i\l\,\hi_\Phi \hi_\Phi} (
\sum C^{(n)}\,e^B)\right]
e^{\l\,F}\right).
\labell{finalcs}
\eeq
Implicitly, eqs.~\reef{finalbi} and \reef{finalcs} employ static gauge
where the two worldsheet coordinates
are identified with two spacetime coordinates. We have chosen
$\tau=t=x^0$ and $\sigma=x^9$. In these expressions,
$P[\cdots]$ denotes the pullback of the enclosed spacetime tensors
to the worldsheet. The $\Phi^i,\ i=1\dots8$, are the transverse scalars,
which are $N\times N$ matrices in the adjoint representation
of the $U(N)$ worldsheet gauge symmetry. The
notation $\hi_\Phi$ denotes the interior product by $\Phi^i$ regarded
as a vector in the transverse space, {\it e.g.}, acting on a two-form
$C^{(2)}={1\over 2}C^{(2)}_{\mu\nu} dx^\mu dx^\nu$, we have
\beq
\hi_\Phi\hi_\Phi C^{(2)}={1\over2}[\Phi^i,\Phi^j]\, C^{(2)}_{ji}
\ .
\eeq
In both eqs.~\reef{finalbi} and \reef{finalcs}, the gauge trace
indicated by $\STr(\cdots)$ is a symmetrized trace. The precise
prescription proposed in ref.~\cite{dielec} was that inside the
trace one takes a symmetrized average over all orderings of the
$F_{ab}$, $D_a\Phi^i$, $i[\Phi^i,\Phi^j]$, and also the individual
$\Phi^i$ appearing in the functional dependence of the background
supergravity fields.
We refer the reader to ref.~\cite{dielec} for more details on these
actions. 

In ref.~\cite{dielec}, the D-particle version of this action was used to
analyze the behavior of $N$ D-particles when placed in a constant
background RR field $F^{(4)}$. This RR four-form is the field strength
associated with D2-brane charge, and ordinarily D0-branes would be
considered neutral with respect to this field.
However, new couplings to the corresponding RR potential $C^{(3)}$
appear in the nonabelian Chern-Simons action (\ref{finalcs}) of the
D-particles. As a result, the D-particles are `polarized' by
the external field into a noncommutative two-sphere, which can
be interpreted
as a spherical D2-D0 bound state. This analysis can readily be
generalized to D$p$-branes in a background of constant
$F^{(p+4)}$. Starting with a flat D$p$-brane with spatial geometry
$R^p$, it will be energetically favorable for the brane to expand into a
non-commutative $R^p\times S^2$ structure. For instance, in the case of
D-strings, the Chern-Simons action (\ref{finalcs}) involves a coupling
\beq
 i\lambda\mu_1\int \Tr P[\hi_\Phi \hi_\Phi C^{(4)}]\ .
\labell{csterm}
\eeq 
Using the same manipulations as in ref.~\cite{dielec}, assuming
$F^{(5)}$ constant (in space), this term produces a new contribution
to the scalar potential of the form 
\beq 
{i\over3}\lambda^2\mu_1\int
dt d\sigma\,\Tr\left(\Phi^i \Phi^j\Phi^k\right)F^{(5)}_{t\sigma
ijk}(t).  
\eeq
As in ref.~\cite{dielec}, one can find solutions for $\Phi^i$ in terms of $N$
dimensional representations of the $SU(2)$ algebra, describing a
D-string with an $R^1\times S^2$ structure. The energy and radius of
this solution can also be calculated from a dual perspective from the
D3-brane action with appropriate world volume gauge fields excited,
corresponding to dissolved D-strings. The two approaches exactly agree
in the large $N$ limit. Similar calculations in the dual 
D3-brane theory appear in ref.~\cite{roberto}.

In this paper we consider similar solutions of the nonabelian
D-string theory. The scalar field configuration again has
a similar interpretation in terms of noncommutative geometry
such that spatial slices have the topology $R\times S^2$. However,
now the
$\Phi^i$ matrices depend on the worldsheet coordinate $\sigma$,
and so the radius of the two-sphere can vary along the length
of the D-string. In
section 3.1, we study possible solutions in a flat space
background, and we obtain configurations corresponding to $N$ D-strings
attached to a D3-brane. We compare features of this solution -- such
as total energy, couplings to background fields, and the shape of the
configuration -- to the corresponding aspects in the D3-brane
spike. In the large $N$ limit, there is an exact correspondence.
Note that there are no nontrivial background fields here, and so these
constructions are quite distinct from the dielectric effect discussed
above and in ref.~\cite{dielec}. 

Following the initial papers on the D3-brane spike \cite{calmald,
gibb, selfd}, there has been a considerable literature studying
various generalizations. This includes the construction of dyonic
spikes describing ($p,q$)-strings
\cite{gauntlett}, analysis of double-funnel solutions
\cite{calmald, gibb, aki}, solutions in
an additional D3-brane supergravity background \cite{gauntlett2,
ghoroku}, and solutions with fundamental strings dissolved in the
D3-brane \cite{khashimoto}. We give a
sample of how these situations can be described from the dual D-string
picture. In section 3.2,  we show the existence of the 
double-funnel solutions describing D-strings stretched between a
D3-brane and an anti-D3-brane (or another D3-brane).
In section 3.3, we give a
brief description of how to construct ($p,q$)-string configurations,
while section 3.4 demonstrates how the BPS funnels survive
even when the system is put in a supergravity background corresponding
to a collection of D3-branes.

A D-string stretched between two D3-branes is represented as a
nonabelian BPS monopole in the D3-brane theory. The `dual' D-string
description provides a physical realization \cite{ded} of the Nahm
equations \cite{nahm}. This interpretation has already received
extensive attention in the literature
\cite{ded,gauntlett,giveon,sethi,hashimoto,gorsky}. In this paper we
consider the primarily the case of the D-string funnel, in which
one of the D3-branes has moved off to infinity.
Our focus is on the large $N$ limit at the point in monopole moduli space
where all the monopoles coincide. This allows us to make a direct
comparison with the bion spikes in the $U(1)$ theory on the
remaining D3-brane. In
particular, using the new  terms in the nonabelian Chern-Simons
action (\ref{finalcs}) \cite{dielec,watiprep}, we can explicitly 
show that the funnel has couplings corresponding to a D3-brane.

The dynamics of the D3-brane spike has also been considered 
\cite{larspeet,fluct,K/T}. In section \ref{fluctsec}, we analyze small
fluctuations propagating along the D-string in the funnel configuration.
 We study both modes that are transverse
and parallel to the D3-brane. Again, in the large $N$ limit, we obtain
exact agreement with the D3-brane analysis \cite{larspeet,fluct,K/T},
in spite of the fact that the present calculation involves noncommuting
matrices and looks rather different. At finite $N$, we find
significant discrepancies with the D3-brane analysis \cite{K/T}
for the higher $\ell$ modes. In particular, due to the noncommutative
character of the funnel, the spectrum is truncated at $\ell_{\rm max}
=N-1$. This suggests a resolution of the puzzle appearing
in ref.~\cite{K/T}, which was mentioned above.

We conclude in Section \ref{discsec} with some further discussion and
comments on our results.


\section{D3-branes from D-strings} \label{d3d1}

In this section we will describe various solutions in the nonabelian
world-volume theory of a D-string corresponding to the D-string
opening up into a D3-brane.

\vskip 1ex
\noindent{\bf 3.1 The BPS Funnel} 

In flat background, the Chern-Simons part (\ref{finalcs}) of the
D-string action plays no role, while the Born-Infeld action (\ref{finalbi})
reduces to \cite{yet2,dielec}\footnote{In these expressions, 
the transverse space indices are raised and lowered with
$g_{ij}=\delta_{ij}$ and $g^{ij}=\delta^{ij}$.} 
\beq
S=-T_1\int d^2\sigma\, \STr\sqrt{-\det\left(\eta_{ab}+
\lambda^2\partial_a\Phi^i Q^{-1}_{ij}\partial_b\Phi^j\right)
\ \det\left(Q^{ij}\right)}\ ,\labell{action2}
\eeq
where
\beq
Q^{ij}=\delta^{ij}+i\lambda[\Phi^i,\Phi^j]\ .
\eeq
Implicitly here, we have set the world-volume gauge field to zero.
This is consistent with the equations of motions for the scalar
field configurations considered here. However, the gauge field will
play an essential role in section 3.3 below. Recall that we are
using static gauge and have chosen
the world-volume coordinates to be $\tau=t=x^0$
and $\sigma=x^9$.
Expanding this action \reef{action2} to leading order (in $\lambda$),
yields the usual nonabelian scalar action
\beq
S\simeq-T_1\int d^2\sigma\, \left( N+{\lambda^2\over2}
\Tr\left(\partial^a\Phi^i \partial_a\Phi^i+
{1\over2}[\Phi^i,\Phi^j][\Phi^j,\Phi^i]\right)+\ldots
\right)
\ .\labell{action3b}
\eeq 
Varying this action yields the following equation of motion
\beq
\prt^a\prt_a\Phi^i=[\Phi^j,[\Phi^j,\Phi^i]]\ .
\labell{motion}
\eeq

Now we are looking for solutions which represent the D-string expanding
into a D3-brane, analogous or `dual' to the bion solutions of the D3-brane
theory \cite{calmald,gibb}. The corresponding geometry would
be a long funnel where the cross-section at fixed $\sigma$
has the topology of a
two-sphere. Hence motivated by the noncommutative two-sphere constructions
of refs.~\cite{wati2,dielec}, we consider the spherically symmetric ansatz
\beq
\Phi^i = \hR(\sigma)\,\alpha^i,\ \ i=1,2,3,\labell{ansatz}
\eeq
where the $\alpha^i$ give some $N\times N$ matrix representation of
the $SU(2)$ algebra
\beq
[\alpha^i,\alpha^j]=2i\,\veps_{ijk}\,\alpha^k\ .
\labell{sutwo}
\eeq
Now at fixed $\sigma$, this ansatz for nonabelian
scalars describes a noncommutative two-sphere with a
physical radius given by
\beq
R(\sigma)^2={\l^2\over N}\sum_{i=1}^3\Tr[\Phi^i(\sigma)^2]=\l^2\,C\,
\hR(\sigma)^2\ .
\labell{radii}
\eeq
Here $C$ is the quadratic Casimir of the particular representation
of the generators under consideration, defined by the identity
\beq
\sum_{i=1}^3(\alpha^i)^2 = C\  {\rm I}_{N}\ ,
\labell{casimir}
\eeq
where ${\rm I}_{N}$ is the $N\times N$ identity matrix. 
For example, $C=N^2-1$
for the irreducible $N\times N$ representation.

Now given the ansatz \reef{ansatz}, the matrix equations of motion
\reef{motion} reduce to a single scalar equation
\beq
\hR''(\sigma)=8 \hR(\sigma)^3\ .\labell{leom}
\eeq
Considering a trial solution, $\hR\propto \sigma^p$, yields
\beq
\hR(\sigma)=\pm {1\over2(\sigma-\signot)}\ , \labell{spike}
\eeq
where we have used the translation invariance of eq.~\reef{leom}
to introduce the integration constant $\signot$. Since
the second order equation \reef{leom}
should have a general solution with two integration constants, it
is clear that eq.~\reef{spike} is not the most general solution
--- we will leave this solution to the next section. However,
this solution \reef{spike} indeed describes the desired funnel,
with the D-string  opening up into a three-brane at $\sigma=\signot$,
where the radius of the funnel diverges.
As it stands eqs.~\reef{ansatz} and \reef{spike} only represent a
solution of the leading order equations of motion \reef{motion},
and so naively one expects that it should only be valid for small
$\hR$ or small radius. However, we will find that this configuration
also solves the full equations of motion extremizing the action
\reef{action2}.

Before plunging into the full equations of motion, let us 
investigate the supersymmetry of the funnel configuration above.
Following the analysis of Callan and
Maldacena \cite{calmald}, we investigate the linearized supersymmetry
conditions, which strictly speaking would only apply for the leading
order action
\reef{action3b}.\footnote{Supersymmetry conditions for the full nonabelian
Born-Infeld action \reef{action2} would be expected to be more
complicated \cite{yet2,rocek}.}
We may write the linearized conditions as
\beq
\Gamma^{\mu\nu}F_{\mu\nu}\,\epsilon=0\ ,
\labell{susy}
\eeq
where $\mu,\nu$ are ten-dimensional indices
and $\epsilon$ is some constant spinor. The latter world-volume
supersymmetry parameter also satisfies the usual D-string
projection \cite{Polchin2}: $\Gamma^{09}\epsilon=\epsilon$. Note that
$\epsilon$ transforms a spinor under both the $SO(1,1)$ Lorentz
transformations of the D-string world-volume theory, and the
$SO(8)$ rotations of the transverse space. Hence it is reasonable
to multiply $\epsilon$ by ten-dimensional Dirac matrices, such
as $\G^{\mu\nu}=[\G^\mu,\G^\nu]/2$.
Following the standard notation (see, \eg refs.~\cite{watirev}),
where $F_{ab}$ denotes the world-volume gauge field strength
which vanishes in the present case, one also has
\beq
F_{ai}=D_a\Phi^i\ , \;\;\;\;\;F_{ij}=i\left[\Phi^i,
\Phi^j\right]\ .
\eeq
Hence eq.~\reef{susy}
yields
\beq
\left(2\Gamma^{\sigma i}D_\sigma \Phi^i+i\Gamma^{jk}\left[\Phi^j,
\Phi^k\right]
\right)\epsilon=0\ .
\eeq
This condition can be solved by spinors satisfying
the projection
\beq
\G^{\sigma 123}\epsilon=\pm\epsilon\ ,
\labell{half}
\eeq
provided that the scalars satisfy the Nahm equations \cite{ded}
\beq
D_\sigma \Phi^i=\pm {i\over2}\varepsilon^{ijk}\left[\Phi^j,
\Phi^k\right]\ .\label{nahmeq}
\eeq
Now inserting our ansatz (\ref{ansatz}) this implies
\beq
\hR'=\mp 2\hR^2\ .\labell{susyeq}
\eeq
However, the solution of this equation is precisely that
given in eq.~\reef{spike}. Hence, we conclude
that the funnel configurations given by eqs.~\reef{ansatz} and
\reef{spike} are in fact BPS solutions preserving $1/2$ of the supersymmetry
of the D-string theory \reef{action3b}. From ref.~\cite{aki}, we
can infer that BPS solutions of the leading order theory \reef{action3b}
are also BPS solutions of the full nonabelian Born-Infeld action
\reef{action2}. That is, the funnel solutions will also solve the
full equations of motion, as we will explicitly demonstrate below. In
a related discussion, ref.~\cite{gauntlett} showed, in the context of
the full nonabelian D-string theory (\ref{action2}), that supersymmetric 
configurations satisfying eq.~\reef{nahmeq} minimize the energy of the system.

We begin by substituting our ansatz (\ref{ansatz}) directly into the 
action (\ref{action2}), and find that it becomes
\beq
S=-T_1\int d^2\sigma\, \STr\sqrt{\left(1+\lambda^2 \alpha^i\alpha^i (\hR')^2
\right)\left(1+4\lambda^2 \alpha^j\alpha^j \hR^4\right)}\ ,\labell{action3}
\eeq
where both $i$ and $j$ are summed over 1,2,3. In deriving this
result, we have eliminated certain combinations of matrices from 
the determinants (and inverses) which will cancel under the
symmetrized trace. In the remaining expression, symmetrization applies to
each of the individual generators $\al^i$ appearing there. Now extremizing
this action \reef{action3} with respect to variations of $\hR$ yields
an equation of motion which may be written as
\beq
{1\over \hR'}
{d\ \over d\sigma}\STr\sqrt{{1+4\lambda^2 \alpha^j\alpha^j \hR^4\over
1+\lambda^2 \alpha^i\alpha^i (\hR')^2}}=0\ .
\labell{motion2}
\eeq
If the radius profile satisfies the supersymmetry
constraint \reef{susyeq}, then the expression under the square root
is simply the identity and it follows that the equation of motion is satisfied.
Hence the supersymmetric funnel solutions are in fact solutions of
the full nonabelian equations of motion \reef{motion2}. Note that
we were able to derive this result without making an expansion
(in $\lambda$) of the matrix expression in eq.~\reef{motion2}
and explicitly implementing the symmetric trace on the individual
terms in this expansion.

It is clear that the funnel solution, eqs.~\reef{ansatz} and
\reef{spike}, describes the nonabelian D-string opening up into a
three-brane on the ($x^1$,$x^2$,$x^3$) hypersurface
at $\s=\signot$. While the
natural intuition is that the latter is actually a D3-brane, it
remains to be demonstrated. We begin by comparing our funnel
solution to the D3-brane monopole or spike \cite{calmald}.
For these purposes, we will focus on the funnel
where the $\alpha^i$ are chosen as the irreducible $N\times N$ 
representation, with $C=N^2-1$. In this case, the radius 
\reef{radii} becomes \cite{giveon}
\beq
R={N\pi \ls^2\over\sigma-\signot}\sqrt{1-1/N^2}\ .\labell{rsigma}
\eeq
To leading order for large $N$, 
this yields precisely (including numerical coefficient) 
with the corresponding formula for the height of D3-brane spike
\cite{calmald}, \ie
\beq
\sigma-\signot={N\pi \ls^2\over R}\ .\labell{reverse}
\eeq
This remarkable agreement is perhaps more than one should
expect, since the D3-brane analysis is strictly speaking only valid for $R$
large, while the current calculations will be reliable for small $R$.
We will comment more on this in the discussion section.

To further corroborate the fact that our funnel yields
a D3-brane, let us compare the energy and some couplings 
to those obtained from the dual D3-brane action. Given our
static solution, the energy is easily derived from the
D-string action (\ref{action3}). Note that using the supersymmetry
condition \reef{susyeq}, the two expressions under the square root
are equal and hence the action is `linearized' \cite{gauntlett}.
We are then left with
\beqa
E&=&T_1\int d\sigma\, \STr\left|1+4\lambda^2 \alpha^i\alpha^i \hR^4\right|
\nonumber\\
&=&2NT_1\int d\sigma\, \hR^2 |\hR'|\left[\left({d\sigma\over d\hR}\right)^2
+\lambda^2 C\right]\ ,
\labell{energy}
\eeqa
where we have repeatedly applied $\hR'=\pm 2\hR^2$ 
in producing the second expression.
We can further manipulate this result by introducing the
physical radius $R=\l\sqrt{C}|\hR|$, as well as using $T_1=4\pi^2 \ls^2 T_3$
to put this expression in the form
\beq
E=T_3{N\over \sqrt{C}}\int4\pi R^2 dR
\left[\left({d\sigma\over dR}\right)^2+1\right]\ .
\labell{energy2}
\eeq
In the dual D3-brane picture, the energy of any (spherically
symmetric) BPS configuration\footnote{Note that this expression
holds regardless of whether electric fields, magnetic fields, or both are
excited on the D3-brane. Hence the agreement found
here is more generally applicable.}
is simply given by \cite{calmald,gauntlett}
\beqa
E&=&T_3\int d^3 x\,[1+(\nabla \sigma)^2]\nonumber\\
&=&T_3\int 4\pi R^2 dR\left[1+\left({d\sigma\over dR}\right)^2\right]\ .
\labell{energy3}
\eeqa
If we chose the irreducible representation,
$N/\sqrt{C}=(1-1/N^2)^{-1/2}$, and hence the energy calculated in these
two formulations agrees up to $N^{-2}$ corrections for large $N$.

If a D3-brane is emerging in the funnel solution, this configuration
should act as a source for the RR four-form potential, $C^{(4)}$.
Such a coupling arises in the Chern-Simons action \reef{finalcs} because of
the nonabelian expectation value of the scalars in this solution.
To leading order, we can focus on the interaction given in \reef{csterm},
which yields
\beqa
i\lambda\mu_1\int \Tr P[\hi_\Phi \hi_\Phi C^{(4)}]
&=&{i\l^2\mu_1\over2}\int d^2\sigma \,C^{(4)}_{tkji}(\tau,\sigma)\,\Tr\left(
\partial_\sigma\Phi^k\,[\Phi^i,\Phi^j]\right)+\ldots
\nonumber\\
&=&\mp i\mu_3{N\over\sqrt{C}}\int dt\,4\pi R^2 dR\ C^{(4)}_{t123}(t,R)\ .
\labell{cstermb}
\eeqa
Here we have used the ansatz \reef{ansatz} and $R=\l\sqrt{C}|\hR|$,
as well as $\mu_1=4\pi^2 \ls^2 \mu_3$ and $\Tr(\al^i\al^j)={N\over3}C\,
\delta^{ij}$. 
In the dual D3-brane formulation, essentially
the same expression arises in
the standard coupling to the RR four-form
\beqa
\mu_3\int P[C^{(4)}]&=&
\mu_3\int dt\,dR\,d\theta\,d\phi\ C^{(4)}_{tijk}\,
\partial_\sigma x^i \partial_\theta x^j \partial_\Phi x^k+\ldots
\nonumber\\
&=&\mu_3\int dt\, 4\pi R^2 dR\,  C^{(4)}_{t123}(t,R)\ .
\labell{csterm3}
\eeqa
So once again if we chose the irreducible representation, we would have
$N/\sqrt{C}=(1-1/N^2)^{-1/2}$, and for large $N$ eqs.~\reef{cstermb}
and \reef{csterm3} agree up to $N^{-2}$ corrections. It is interesting
that in deriving this agreement for
the RR coupling, we only used the basic ansatz \reef{ansatz}, but not
the details of the funnel solution \reef{spike}. Hence this result
will hold more generally, and in particular it still holds in the
following sections.
In eq.~\reef{cstermb}, the minus (plus) sign arises if $\hR$ is
positive (negative). Hence
this calculation shows that the minus-solution in eq.~\reef{spike}
corresponds to the D-string opening up into a D3-brane
(assuming we approach from $\sigma>\signot$), while
the plus-solution has the opposite orientation and corresponds to an
anti-D3-brane.

To summarize this section, we have shown that by allowing for
suitable boundary conditions in the nonabelian D-string theory,
the latter can `grow' into a D3-brane. This construction is a dual
formulation of the BPS magnetic monopole in the abelian D3-brane theory
which describes a D-string spike growing out of the three-brane surface.
In the present calculation, we see that the geometry at the
core of the spike is noncommutative, with the level of discreteness set by $N$,
the number of D-strings. 
In these last few calculations, we have focused on using the
irreducible $N\times N$ representation of the $SU(2)$ generators
\reef{sutwo}, and we found good quantitative agreement at large $N$
between the two formulations. These calculations indicate that
the funnel solution describes the $N$ D-strings expanding into
a single fundamental D3-brane. Using reducible representations
would correspond to creating several (independent) D3-branes from
the same $N$ D-strings. Paralleling the constructions in
ref.~\cite{dielec}, one could then construct
multi-center funnels located at different positions in the
($x^1,x^2,x^3$) hypersurface.

\vskip 1ex
\noindent{\bf 3.2 Double Funnels} 

With the ansatz \reef{ansatz}, the leading-order matrix equations
became $\hR''=8\hR^3$ in eq.~\reef{leom}. As a first step to generating
the most general solution, we integrate this equation as
\beq
(\hR')^2=4(\hR^4-\hR^4_0)\ ,
\labell{first}
\eeq
where $\hR^4_0$ is an arbitrary integration constant. In principle then,
integrating once more yields the general solution 
\beq
\sigma=\signot\pm{1\over2}\int^\infty_\hR {d\tR\over\sqrt{\tR^4-\tR^4_0}}\ .
\labell{second}
\eeq
This solution looks remarkably similar to those describing double
funnels or wormholes in dual D3-brane framework \cite{calmald,gibb}.

However, before examining the details of these configurations, let us
consider the analogous solutions of the full equation of motion
\reef{motion2}. The symmetrized prescription \cite{yet,dielec}
instructs us to expand the square root expression and symmetrize over all
permutations of the generators $\alpha_i$ in the trace of each term
in the expansion. For example:
$\STr(\alpha^i\alpha^i)=N C$,
$\STr(\alpha^i\alpha^i\alpha^j\alpha^j)=N (C^2-4C/3)$, and
$\STr(\alpha^i\alpha^i\alpha^j\alpha^j\alpha^k\al^k)=N (C^3-4C^2+16C/3)$.
Unfortunately, we have not been able to find a systematic construction
for the general term in this expansion. However, observing
that at leading order, $\STr(\alpha^i\alpha^i)^m\simeq N C^m$, we can
construct an approximate equation by replacing the $\alpha^i\alpha^i$ by
$C\,{\rm I}_N$ in eq.~\reef{motion2}. For large $N$, this keeps the
leading order contribution at every order (in $\l$) in the expansion
of the square roots. Within this approximation the equation of motion
becomes
\beq
{N\over \hR'}\,
{d\ \over d\sigma}\sqrt{{1+4\lambda^2 C \hR^4\over
1+\lambda^2 C (\hR')^2}}=0\ .
\labell{motion2b}
\eeq
Integrating this equation is trivial, and the result may be 
expressed as
\beq
(\hR')^2=4{\hR^4-\hR_0^4\over1+4 \l^2C \hR_0^4}\ .\labell{solu}
\eeq
In terms of the physical radius \reef{radii}, we have
\beq
(R')^2={4}{R^4-R_0^4\over\l^2C+{4R_0^4} }\ ,\labell{solu2}
\eeq
where we have also rescaled the integration constant in the obvious way.
The solution of this equation is then implicitly given by
\beq
\sigma=\signot+{1\over2}\int^\infty_R d\tR\  
\sqrt{\l^2 C+{4R_0^4}\over \tR^4-R_0^4}\ .
\labell{solu3}
\eeq

With $C=N^2-1$ for the irreducible representation, eq.~\reef{solu3}
precisely reproduces the general solutions constructed in
refs.~\cite{calmald,gibb} for large $N$.
For $R_0=0$ we recover the supersymmetric
funnel solution \reef{rsigma}. For large $R$, the general solution
approximates this funnel and so given our previous discussion the
nonabelian D-string is again expanding into a D3-brane at $\sigma=\signot$.
Assuming $R_0^4>0$, we see from eq.~\reef{solu2} that the noncommutative
funnel stops contracting when $R=R_0$. The obvious solution to
continue past this point is \cite{calmald,gibb}
\beq
\sigma=\signot+2\Delta\sigma-{1\over2}\int^\infty_R d\tR\ 
\sqrt{\l^2 C+{4R_0^4}\over R^4-R_0^4}\ ,
\labell{solu3b}
\eeq
where $\Delta\sigma=\sigma(R_0)-\signot$. Hence beyond the minimum
radius, the solution re-expands into an anti-D3-brane at $\sigma=
\signot+2\Delta\sigma$.\footnote{Verifying that the emergence
of a  D3-brane or an anti-D3-brane at either end of these double
funnel solutions actually requires examining the sign of $\hR$ as
$\sigma =\signot$ and $\signot+2\Delta\sigma$, as per the discussion
in the section 3.1.}
Thus we have reproduced the wormhole solutions
of refs.~\cite{calmald,gibb}
from the point of view of the nonabelian D-string theory.

In integrating eq.~\reef{motion2b}, one could also choose a
negative integration constant, in which case it is natural to
write
\beq
(R')^2=(R_0')^2+4{1+(R_0')^2\over\l^2C}R^4\ ,\labell{solu4}
\eeq
where $R_0'$ is a new dimensionless integration constant.
The general solution then becomes
\beq
\sigma=\signot+{1\over2}\int^\infty_R 
{\l\sqrt{C}\,d\tR\over\sqrt{(1+(R_0')^2) \tR^4+\l^2C(R_0')^2/4}}\ .
\labell{solu5}
\eeq
In this case, the funnel collapses all the way down to zero radius,
which is approached with a finite slope, \ie from eq.~\reef{solu4},
$R'(R=0)=R_0'$. The integrand has no singularity at $\tR=0$,
and so one can continue the solution beyond this point if one
allows the radius to become negative. Alternatively, keeping the
radius positive, we would match eq.~\reef{solu5} onto 
\beq
\sigma=\signot+2\Delta\sigma-{1\over2}\int^\infty_R
{\l\sqrt{C}\,d\tR\over\sqrt{(1+(R_0')^2) \tR^4+\l^2C(R_0')^2/4}}\ .
\labell{solu5b}
\eeq
where now $\Delta\sigma=\sigma(R=0)-\signot$. Hence in this solution,
the funnel collapses down to zero size and then re-expands
into another D3-brane$^4$ at $\sigma=\signot+2\Delta\sigma$.
These general solutions again match, for large $N$, the analogous cusp
configurations constructed in the D3-brane framework \cite{aki}.
For comparison purposes, it may be simpler to think of these solutions
in the form given in eqs.~\reef{solu3} and \reef{solu3b}, but with
$R_0^4<0$. Note that in this case, we must choose $-\l^2C/4<R_0^4$
to produce a real solution. This lower bound $R_0^4\rightarrow
-\l^2C/4$ corresponds to the singular limit $R_0'\rightarrow \infty$.

These cusp solutions describe $N$ D-strings stretched between two parallel
D3-branes (or anti-D3-branes),
which should be a supersymmetric configuration. However, the supersymmetry
condition \reef{susyeq} is only satisfied when $R_0'=0$, in which case
the D-string extends off to infinity before reaching zero size.
In the dual D3-brane framework, Hashimoto \cite{aki} identified the
correct supersymmetric solution for $N=1$
as a BPS monopole of the nonabelian
world-volume theory describing the two D3-branes. To find the
corresponding BPS solutions in the D-string theory, one must begin with an
ansatz more general than eq. (\ref{ansatz}) when solving the Nahm
equations (\ref{nahmeq}). Such solutions are known, see for example
refs.\ \cite{prasad, hashimoto}. We leave a discussion of these solutions
for future work \cite{son}.

\vskip 1ex
\noindent{\bf 3.3 ($p,q$)-Strings} 

The previous analysis is readily generalized to
($p,q$)-strings, \ie bound states of D-strings and fundamental strings
\cite{pqsting}.
This is done by simply introducing a background $U(1)$
electric field on the D-strings, corresponding to fundamental
strings dissolved on the worldsheet. 
Denoting the electric field as $F_{\tau\s}=\E\,{\rm I}_N$, the D-string action
(\ref{action3}) becomes
\beq
S=-T_1\int d^2\sigma\, \STr\sqrt{\left(1-\lambda^2 \E^2+\lambda^2 \alpha^i\al^i
(\hR')^2\right)\left(1+4\lambda^2 \alpha^j\alpha^j \hR^4\right)}\ ,
\labell{action9}
\eeq
where we have inserted the noncommutative ansatz \reef{ansatz}.
Extremizing with respect to variations of $\hR$ yields
\beq
{1\over \hR'}
{d\ \over d\sigma}\STr\sqrt{{1+4\lambda^2 \alpha^j\alpha^j \hR^4\over
1-\lambda^2 \E^2+\lambda^2 \alpha^i\alpha^i (\hR')^2}}=0\ .
\labell{motion2z}
\eeq
Now assuming $\E$ constant, a simple rescaling of eq.~\reef{susyeq}
yields an exact solution, \ie
\beq
\hR'=\pm2\sqrt{1-\lambda^2 \E^2}\,\hR^2\ .
\labell{neweq}
\eeq
Hence the funnel solution for the ($p,q$)-string becomes
\beq
\hR(\sigma)={1\over2\sqrt{1-\lambda^2 \E^2}}{1\over\s-\signot}\ .
\labell{newspike}
\eeq
(For simplicity, we will only consider the positive root of eq.~\reef{neweq}
in the following.)
It is also useful to consider the electric displacement $D$, conjugate to $E$,
\beq
D\equiv{1\over N}{\delta S\over\delta\E}={1\over N}\STr\sqrt{1+4\lambda^2
\al^i\al^i \hR^4\over 1-\lambda^2 \E^2+\lambda^2 \al^j\al^j
(\hR')^2}\ 
 \lambda^2 T_1 \E={\lambda^2 T_1 \E\over\sqrt{1-\lambda^2 \E^2}}
\labell{displace}
\eeq
where we have used eq.~\reef{neweq} to derive the final result.
One can verify that the equations of motion for the world-volume gauge field
specify $D$ to be a constant, and so 
our assumption of constant $\E$ is consistent for any solution
$\hR(\sigma)$ obeying eq.~\reef{motion2z}. 
For $N_f$ fundamental strings, one obtains the correct $(p,q)$-string
tension by quantizing $D=N_f/N$, remembering that the fundamental
string tension is simply $1/\lambda$ --- see below.

To determine the energy of the system, we must evaluate the Hamiltonian,
$\int d\sigma(D\E-\cal{L})$, for the dyonic funnel solutions.
Manipulating this expression in a manner similar to the analogous
calculations in eqs.~(\ref{energy},\ref{energy2}), the final result may be 
expressed as a sum of two terms
\beq
E=T_1\int d\sigma \sqrt{N^2+g^2 N_f^2}+T_3{N\over\sqrt{C}}
\int 4\pi R^2dR\ .
\labell{newenergy}
\eeq
where $g$ is the string coupling, and we remind the reader that
$T_1=(\l g)^{-1}$. Here the first term comes from collecting the
contributions independent of $\hR$, and correctly matches the energy of
the $(N,N_f)$-string bound state \cite{pqsting}. The second
contribution involves the terms containing $\hR$ and as in section
3.1, eq.~\reef{neweq} is used to put these in the form $\hR^2\, |\hR'|$.
The final result
corresponds to the expected energy of an orthogonal D3-brane, at large $N$.
Our expression \reef{newenergy} also matches the expectations (for large $N$)
from the similar calculations in the D3-brane theory \cite{gauntlett}. 
Eq.~\reef{energy3} still applies in this case, and in this
formulation the $(\nabla\sigma)^2$ term provides the contribution
of the $(N,N_f)$-string. In the D3-brane formulation, it is straightforward
to show that the dyonic spike is still supersymmetric. However, in the
D-string formulation, introducing a constant background electric field
moves the theory to a new superselection sector where the supersymmetry
is nonlinearly realized.\footnote{We would like to thank Amanda Peet
for a discussion on this point.} By the $SL(2,Z)$ duality of the
type IIb superstring theory, it is clear that a $(N,N_f)$-string
has precisely the same amount of supersymmetry as an ordinary
D-string. Similarly the dyonic funnel will be a BPS configuration
preserving $1/2$ of the world-volume supersymmetries. This supersymmetry
is reflected in that
these configurations \reef{neweq} satisfy the full equations of motion
\reef{motion2z}, and that the corresponding energy \reef{newenergy}
splits into a sum of string and three-brane contributions.

There are
also various other solutions known in the literature
involving ($p,q$)-strings (see, for example, ref.~\cite{ghoroku}), and
these should presumably also follow from the full equations of motion
\reef{motion2z} in a straightforward manner. For related discussion on
string junctions, see refs.~\cite{hashimoto,gorsky}.

\vskip 1ex
\noindent{\bf 3.4 Embedding in a D3-brane Background} 

We now consider constructing a noncommutative funnel for a 
nonabelian D-string sitting
in the background of a set of orthogonal D3-branes. According
to refs.~\cite{gauntlett2,ghoroku}, when working with a test
D3-brane sitting in such a supergravity background, the BPS Born-Infeld
spike solutions are unchanged by the background fields. Hence one
might expect that the funnel solutions (\ref{ansatz},\ref{spike}) will also
appear unchanged in the modified world-volume theory of the D-strings.

An extremal D3-brane background can be written as\cite{solute}
\beqa
ds^2&=&{-dt^2+(dx^i)^2\over\sqrt\cH}+\sqrt{\cH}(dx^m)^2
\nonumber\\
F^{(5)}&=&\mp\cH^{-2}\prt_m\cH\,dt\wdg dx^1\wdg dx^2\wdg dx^3\wdg dx^m
\labell{back}\\
&&\ \ \pm\prt_m\cH\,i_{\hat{x}^m}\left(dx^4\wdg dx^5\wdg dx^6\wdg dx^7
\wdg dx^8\wdg dx^9\right)
\nonumber
\eeqa
where the $x^i,\ i=1\ldots3$, 
directions are parallel to the D3-brane, and the $x^m,\ m=4\ldots9$,
directions are transverse. The function $\cH$ satisfies the Laplace
equation: $\prt^m\prt_m\cH=0$. The single center harmonic function is
\beq
\cH=1+4\pi gN_3\left({\ls\over r}\right)^4
\labell{harmonic}
\eeq
for $N_3$ D3-branes, where $r^2=\sum_{m=4}^9(x^m)^2$.
The potential for the five-form field strength has an electric
component which may be written
\beq
\Cf_{elec}=\pm\left(\cH^{-1}-1\right)\,dt\,dx^1\,dx^2\,dx^3\ .
\labell{potential}
\eeq
The magnetic part of the potential will only involve indices in the transverse
$x^m$ directions, and we will argue below that it is irrelevant for the
present calculation.

Now consider $N$ D-strings extending into the transverse space
along the $x^9$-axis. We again choose static gauge for the D-string
action with $\tau=t$ and $\sigma=x^9$. In the nonabelian
Chern-Simons action (\ref{finalcs}), we have the interactions
\beqa
\mu_1\int\, i\lambda\,\STr\,P\left[i_\Phi i_\Phi \Cf\right]
&=&\ {i\over2}\lambda\mu_1\int d^2\sigma\,\STr\left(\Cf_{t9ji}
\,[\Phi^i,\Phi^j]
+\lam \Cf_{tkji}\,{D_\sigma\Phi^k}\,[\Phi^i,\Phi^j]\right.
\labell{csint}\\
&&\qquad\quad\left.
-\lam \Cf_{9kji}\,{D_t\Phi^k}\,[\Phi^i,\Phi^j]
+{\lam^2\over2} \Cf_{l kji}\,{D_t\Phi^l}\,{D_\sigma\Phi^k}
\,[\Phi^i,\Phi^j]\right)\ .
\nonumber
\eeqa
In the solution which we will construct, we assume that:
(i) there are no background gauge fields so the covariant derivatives
in the pullbacks are simply ordinary partial derivatives,
(ii) the solution is static so that $D_t\Phi^l=\prt_t\Phi^l=0$,
and (iii) for later purposes, that only the $\Phi^i$ in the $x^i$
directions are relevant, \ie we only consider ``deformations''
of the D-string in its transverse directions which are parallel
to the world-volume directions of the background D3-brane.
Further, from eqs.~\reef{back} or \reef{potential}, we know that
$\Cf_{t9ji}=0$. Hence the only relevant interaction above is
\beq
{i\over2}\lambda^2\mu_1\int d^2\sigma
\,\STr\left(\Cf_{tkji}\,{\prt_\sigma\Phi^k}\,[\Phi^i,\Phi^j]\right)\ .
\labell{csintb}
\eeq

In the background RR potential and the metric, we have the harmonic function
\reef{harmonic} which is only a function of the nonabelian radius
\beq
r^2=\sigma^2+{\lambda^2}\left[(\Phi^4)^2+(\Phi^5)^2+
\ldots\right]
\simeq \sigma^2\ .
\labell{radius}
\eeq
In the last step, in keeping with the assumptions listed above, 
we have ignored the fluctuations of the D-string in the $x^m$-directions.
This simplifies the calculation since we have $\cH=\cH(\sigma)$, however,
we might expect 
some smearing of the D-string in the $x^m$-directions at higher
order. With this simplification,
the Chern-Simons interaction becomes
\beq
{i\over2}\lambda^2\mu_1\int d^2\sigma
\,\left(\cH(\sigma)^{-1}-1\right)\veps_{kji}\,
\Tr\left({\prt_\sigma\Phi^k}\,[\Phi^i,\Phi^j]\right)\ ,
\labell{csintc}
\eeq
where for definiteness, we have chosen the plus sign for the potential
in eq.~\reef{potential}. This choice corresponds to a background
of D3-branes (as opposed to anti-D3-branes).

The Born-Infeld part of the action is only slightly modified by the
background metric. For our usual ansatz (\ref{ansatz}), the
Born-Infeld action now reads
\beq
S_{\rm BI}=-T_1\int d^2\sigma\,\STr\sqrt{\left(1+{\lambda^2 \over \cH}\al^i\al^i
(\hR')^2\right)\left(1+{4 \lambda^2 \over \cH}\al^j\al^j\hR^4\right)}\ .
\labell{actone}
\eeq
Similarly inserting eq.~\reef{ansatz} into the Chern-Simons interaction
\reef{csintc} yields
\beqa
S_{\rm CS}&=&2\l^2NCT_1\int d^2\s\,\left(\cH(\sigma)^{-1}-1\right)\hR^2\hR'
\nonumber\\
&=&-{2\over3}\l^2NCT_1\int d^2\s\,\prt_\s\left(\cH(\sigma)^{-1}\right)
\hR^3\ .
\labell{acttwo}
\eeqa
Since the function $\cH$ depends on $\sigma$, the full equations of
motion following from this action are considerably more complicated
than in the flat space case. The full equations of motion may be written
as
\beqa
&&-{1\over \hR'}
{d\ \over d\sigma}\STr\sqrt{{1+{4\lambda^2\over\cH} \alpha^j\alpha^j \hR^4\over
1+{\lambda^2\over\cH} \alpha^i\alpha^i (\hR')^2}}
+{1\over \hR'}\STr\left[\left(1+{\lambda^2\over\cH} \alpha^i\alpha^i
(\hR')^2\right){\tilde{d}\ \over \tilde{d}\sigma}\sqrt{{1+{4\lambda^2\over\cH}
\alpha^j\alpha^j \hR^4\over
1+{\lambda^2\over\cH} \alpha^i\alpha^i (\hR')^2}}\right]
\nonumber\\
&&\qquad+\STr\left[\al^k\al^k\sqrt{{1+{4\lambda^2\over\cH} \alpha^j\alpha^j
\hR^4\over 1+{\lambda^2\over\cH} \alpha^i\alpha^i (\hR')^2}}\right]
\l^2\prt_\s\left(\cH(\sigma)^{-1}\right)\hR'
={2}\l^2NC\,\prt_\s\left(\cH(\sigma)^{-1}\right)\hR^2\ .
\labell{beast}
\eeqa
In the second term on the left hand side, $\tilde{d}/\tilde{d}\s$
denotes that the $\s$ derivative only acts on the harmonic
function $\cH$. 
The right hand side of this equation is the contribution from the
Chern-Simons term in eq.~\reef{acttwo}. Now if as in eq.~\reef{susyeq},
we set $(\hR')^2=4\hR^4$, the first two terms vanish and the entire
expression reduces to simply
\beq
\hR'=2\hR^2\ ,
\labell{beauty}
\eeq
with the standard solution
\beq
\hR=-{1\over2(\s-\signot)}\ .
\labell{beauty2}
\eeq
Hence the background picks out the noncommutative funnel which
corresponds to the D-string expanding into a D3-brane, but not
the one where an anti-D3-brane emerges. This should have been expected
because an anti-D3-brane would be unstable in the D3-brane background.
Choosing the opposite sign of the RR potential in eq.~\reef{csintc}
would correspond to putting the D-string in the supergravity background
generated by a collection of anti-D3-branes. This would also change the
sign of the Chern-Simons contribution in the equation of motion
\reef{beast} to produce $\hR'=-2\hR^2$ in place of eq.~\reef{beauty}.
Hence in this case, the noncommutative funnel corresponding to
an anti-D3-brane would be picked out. In any event, the BPS funnel solution
consistent with the supersymmetry of the background survives
unchanged, just as for the analogous solutions found in the
D3-brane formulation. As a final note here, we observe that 
this configuration solves the full equations of motion
\reef{beast} {\it regardless} of the detailed functional form of $\cH$.
In particular, the position of the end of the funnel, \ie $\s=\signot$,
is still an independent parameter, not correlated to the position(s)
of the background D3-branes. Further, one could consider 
multi-center solutions for $\cH$.

\section{Fluctuations of the D-string Funnel} \label{fluctsec}

In this section, we analyze the dynamics of the BPS funnel solution 
(\ref{spike}) (in a flat background).
That is, we examine the linearized equations of motion for
 small, time-dependent fluctuations of the transverse scalars 
$\Phi^r$, around the exact background
$\Phi^i={1\over2\sigma}\alpha^i$.\footnote{For simplicity, we have
set $\signot=0$. We will also
choose the generators for the background solution to lie
in $N\times N$ irreducible representation. Hence $C=N^2-1$.} 
There are two types of fluctuations to consider: The first,
in the language of ref.~\cite{K/T}, 
are the `overall transverse' excitations 
given by the scalars $\delta\Phi^m$ which are transverse 
to both the D-string 
and the noncommutative two-sphere (or D3-brane).
The second are the `relative transverse'
fluctuations of the coordinate fields $\Phi^i$ which lie in 
the two-sphere directions. Our notation in the following
will be that indices $i,j=1\ldots3$ denote the directions
parallel to the D3-brane, $m,n=4\ldots8$ represent directions 
transverse to both the D-string and the D3-brane, and finally
$r,s=1\ldots8$ include all of these directions. 

We start with the overall transverse fluctuations. 
The simplest type of
fluctuation is just proportional to the identity matrix, say $\delta
\Phi^m(\sigma,t) = f^m(\sigma,t){\rm I}_{N}$. For these modes it is
straightforward to plug into the action (\ref{action2}), and we find
\beqa
S&=&-T_1\int d^2\sigma\STr\sqrt{\left(1+\frac{\lambda^2}{4\sigma^4}\al^i\al^i
\right)\left[ \left(1+\frac{\lambda^2}{4\sigma^4}\al^j\al^j\right)\left(1-
\lambda^2(\partial_t\delta\Phi^m)^2\right)
+\lambda^2(\partial_{\sigma}\delta\Phi^m)^2\right]}
\nonumber\\
&\simeq&-NT_1\int d^2\sigma\left[H -{\lambda^2\over2}H(\partial_tf^m)^2
+{\lambda^2\over2}(\partial_{\sigma}f^m)^2+\ldots\right]\ ,
\labell{quada}
\eeqa
where we introduced
\beq
H(\sigma)=1+\frac{\lambda^2C}{4\sigma^4}\ .
\labell{oldh}
\eeq
In the final action, we have only kept the terms quadratic in the fluctuations
as this is sufficient to determine the linearized equations of motion:
\beq
\left(H\partial_t^2-\partial_{\sigma}^2\right)f^m=0\ .
\labell{oteom}
\eeq
This is precisely the equation of motion found for the transverse
fluctuations of the DBI spike soliton in
refs.~\cite{calmald,larspeet,K/T}. The identification of the function $H$
with the corresponding functions in refs.~\cite{calmald,larspeet,K/T}
requires $\lambda^2C/4=\pi^2 N^2\ls^2$, which again holds up to
$1/N^2$ corrections for large $N$.
Note that this equation was also found to agree
with the equation of motion for a fluctuating test string in the 
supergravity background of a D3-brane~\cite{larspeet}, after
identifying parameters on both sides in a specific 
way.\footnote{More precisely, the functional forms of the equations agree,
while the parameters undergo some form of renormalization between the
two pictures. Exact matching occurs only at a specific point in
parameter space, on the border between the two regimes of validity.}

In the detailed analysis of ref.~\cite{K/T}, the fluctuations
in eq.~\reef{oteom} correspond to the $\ell=0$ modes, \ie modes constant
on the two-sphere. 
We will now show that, up to an important modification, 
similar agreement also holds for the higher $\ell$ modes. 
To describe the $\ell>0$ modes we first note, 
following refs.~\cite{wati2,fuzzball}, that the fluctuations 
$\delta\Phi^m$
can be expanded on the non-commutative two-sphere as a polynomial series in the
matrices $\alpha^i$ as follows,
\beq
\delta\Phi^m(\s,t)=\sum_{\ell=0}^{N-1}
\psi^m_{i_1i_2...i_{\ell}}(\s,t)\,\alpha^{i_1}\alpha^{i_2}
\cdots\alpha^{i_{\ell}}
\labell{expand}
\eeq
where the coefficients $\psi^m_{i_1\cdots i_{\ell}}$ are completely symmetric
and traceless in the lower indices. Also note that the series must terminate
after $N-1$ terms since
there are at most this many linearly independent matrices which can be formed
from an $N\times N$ irreducible representation of the $\alpha^i$. In the large
$N$ limit this expansion is analogous to expanding the fluctuations in 
spherical harmonics on a commutative two-sphere. 

Substituting this form of the fluctuations into the action (\ref{action2})
is now slightly more involved, and it is more straightforward to use
an alternative form of the action, given in terms of eq.~(26) in
ref.~\cite{dielec}. In flat space, this form of the action
(\ref{action2}) reads $S=-T_1\int\sqrt{-\tilde{D}}$, with
\beq
\tilde{D}=\det\pmatrix{\eta_{ab}&
\lambda\partial_a\Phi^s\cr
-\lambda\partial_b\Phi^r&\delta^{rs}+i\lambda\,[\Phi^r,\Phi^s]\cr}\ ,
\labell{transform}
\eeq
where $\Phi^r$ includes the background $\Phi^i$ and the overall transverse
fluctuations $\delta\Phi^m$. In general, this leads to a $10\times10$
determinant, which, however, is straightforward to
evaluate keeping in mind the symmetrization procedure. We find
that the resulting action is same as in the first line of eq.~\reef{quada}
up to the addition of an extra commutator term $\lambda^2
[\Phi^i,\delta\Phi^m][\delta\Phi^m,\Phi^i]$ in the second factor
under the square root. The quadratic action then becomes
\beqa
S&\simeq&-T_1\int d^2\sigma\,\Tr\left(H -{\lambda^2\over2}
H(\partial_t\delta\Phi^m)^2
+{\lambda^2\over2}(\partial_{\sigma}\delta\Phi^m)^2
+{\lambda^2\over2}[\Phi^i,\delta\Phi^m][\delta\Phi^m,\Phi^i]\right.
\nonumber\\
&&\qquad\qquad\qquad\left.
+{\lambda^4\over12}[\prt_\s\Phi^i,\prt_t\delta\Phi^m][\prt_t\delta\Phi^m,
\prt_\s\Phi^i]
\right]\ ,
\labell{quadb}
\eeqa
where the last term arises from taking care to expand the symmetrized
trace in the kinetic term.
Now the linearized equation of motion becomes
\beq
\left(H\partial_t^2-\partial_{\sigma}^2\right)\delta\Phi^m 
+\left[\Phi^i,[\Phi^i,\delta\Phi^m]\right]
-{\l^2\over6}\left[\prt_\s\Phi^i,[\prt_\s\Phi^i,\prt_t^2\delta\Phi^m]\right]=0
\labell{fluceqn}
\eeq
In order to make contact with the discussion in ref.~\cite{K/T}, we must 
evaluate the commutator terms in this equation for the background solution
$\Phi^i=\frac{1}{2\sigma}\alpha^i$. To facilitate this we will make use of 
the expansion in eq.~\reef{expand}. Specifically, we evaluate
\beqa
\left[\alpha^i,[\alpha^i,\delta\Phi^m]\right]&=&
\sum_{\ell <N}\psi^m_{i_1i_2\cdots
i_{\ell}}\left[\alpha^i,[\alpha^i,\alpha^{i_1}
\alpha^{i_2}\cdots \alpha^{i_{\ell}}]\right]
\nonumber\\
&=&\sum_{\ell <N}4\ell(\ell+1)\,\psi^m_{i_1i_2\cdots
i_{\ell}}\alpha^{i_1}
\alpha^{i_2}\cdots \alpha^{i_{\ell}}\ ,
\labell{comm}
\eeqa
making use of the fact that $\psi^m_{i_1i_2\cdots i_{\ell}}$ is completely 
symmetric and traceless. So we see that the double commutator above
essentially
acts like the Laplacian on the noncommutative two-sphere.
Hence restricting the fluctuation to contain products with a fixed
number of generators, \ie to contain a specific spherical harmonic
on the two-sphere, we have
\beq 
\left[\Phi^i,[\Phi^i,\delta\Phi_{\ell}^m]\right]=\frac{\ell (\ell 
+1)}{\sigma^2}\delta\Phi_{\ell}^m
\qquad{\rm and}\qquad
\left[\prt_\s\Phi^i,[\prt_\s\Phi^i,\prt_t^2\delta\Phi_{\ell}^m]\right]
=\frac{\ell (\ell 
+1)}{\sigma^4}\,\prt_t^2\delta\Phi_{\ell}^m\ .
\labell{ell}
\eeq
Thus we see that the equation of motion for each mode becomes
\beq
\left(\tilde{H}_\ell\partial_t^2-\partial_{\sigma}^2\right)\delta\Phi_{\ell}^m
+\frac{\ell (\ell +1)}{\sigma^2}\delta\Phi_{\ell}^m\;\;=0
\labell{bigeqn}
\eeq
where
\beq
\tilde{H}_\ell(\sigma)=1+\frac{\lambda^2}{4\sigma^4}
\left(C-{2\over3}\ell(\ell+1)\right)\ .
\labell{newH}
\eeq

Again for large $N$ and $\ell\ll N$, this reproduces
the equation of motion for overall transverse fluctuations
found from the D3-brane spike \cite{K/T}.
However, for large $\ell$ the coefficient in $\tilde{H}_\ell$ is
significantly modified compared to the D3-brane analysis. Another
important difference is that in the noncommutative D-string
analysis, the spectrum of modes is truncated at $\ell_{\rm
max}=N-1$. Note that given this truncation, the coefficient of
$1/\sigma^4$ in (\ref{newH}) can not be negative -- such a negative
coefficient would have caused drastic changes in the mode propagation.
Thus for finite $N$, there are a finite number of modes propagating
in the core of the bion. Note that this number is $\sum_{\ell=0}^{N-1}
(2\ell+1)=N^2$. Of course, this counting is precisely what is
expected for the adjoint scalars in the $U(N)$ theory on the
world-volume of the $N$ D-strings. A puzzle was raised in
the D3-brane analysis \cite{K/T}, where it appeared that
modes with arbitrarily high $\ell$ would propagate
out along the D-string spike. The present D-string analysis
suggests that this is not the case, and that the propagation
of high $\ell$ modes is significantly modified. We will comment
more on this point in the discussion section.

We finally briefly discuss the case of relative transverse fluctuations,
$\delta \Phi^i$. From the
D3-brane point of view these are considerably more complicated to
analyze, because of the interplay between the scalar field and
the gauge field. In the D-string picture, an increased complication
arises in evaluating the symmetrized trace. Unfortunately, we
do not have an exact treatment of the quadratic action. Instead
we use the same approximation as in section 3.2, replacing
in the action everywhere $\al^i\al^i$ by $C\,{\rm I}_N$. For large
$N$, this keeps the leading contribution at every order in $\l$
in an expansion of the action. As above, first let us consider
the $\ell=0$ mode, and for
concreteness consider a fluctuation in the $x^3$-direction, $\delta
\Phi^3(\sigma,t)=f(\sigma,t) {\rm I}_{N}$. One can 
show that fluctuations in different directions decouple at linear
order, using $\Tr(\alpha^i\alpha^j)=NC/3\,\delta^{ij}$. The determinant
in (\ref{transform}) now involves a $5\times5$ matrix which again is
straightforward to calculate. We find
\beq
S=-T_1\int d^2\sigma\,\STr\sqrt{H\left(H-\dot{f}^2+{1\over H}
[1+\alpha_3^2/(4\sigma^4)](f')^2+\alpha_3 f'/\sigma^2\right)}\ ,
\eeq
where $H$ is given in eq.~\reef{oldh}.
Using $\Tr\,\alpha_3=0$ in an expansion in the amplitude $f$,
the terms nicely arrange into
\beq
S=-NT_1\int d^2\sigma\left[H-{\dot{f}^2\over2}+{(f')^2\over2H}+{\cal
O}(f^4)\right].
\eeq
The equations of motion follow immediately, and again agree, for large
$N$, with the results from the D3-brane spike and from supergravity
\cite{larspeet,K/T}.

Higher $\ell$ modes can be treated similarly. For instance, the
$\ell=1$ `breathing' mode considered in ref.~\cite{K/T} is implemented
by the fluctuations $\delta \Phi^i(\sigma,t)=f(\sigma,t)\alpha^i$. The
resulting action is found most easily by substituting $\hR\rightarrow
\hR+f$ in the Born-Infeld action (\ref{action3}), adding the obvious
term involving time-derivatives. The resulting equation of motion
reads
\beq
\partial^2_t f-\partial_\sigma\left({\partial_\sigma f\over H}\right)+
{2\over H^2\sigma^2}(4-H)f=0.
\eeq
The corresponding equation is only given indirectly in
ref.~\cite{K/T}, but through a bit of algebra one can verify that the
two approaches agree, again up to $1/N^2$ corrections. 

\section{Discussion} \label{discsec}

Making use of the recently proposed nonabelian 
extension of the Born-Infeld action describing the world-volume physics 
of D$p$-branes \cite{dielec,watiprep}, we have found a description of a
D-string  ending on a D3-brane `dual' to that obtained from the 
abelian D3-brane action. Specifically
we have shown that the world-volume action for $N$ coincident D-strings 
has {\it exact} BPS solutions which describe a D3-brane growing out of the 
noncommuting transverse coordinates of the collection of D-strings. 
We generalized this construction to describe $(p,q)$-strings by
considering the case where there are a number of fundamental strings
dissolved on the worldsheet of the D-string. We have also considered the 
case where the D-string is embedded in the supergravity background
of a collection of D3-branes, and shown that the supersymmetric
funnel remains a solution with its form unchanged.
In all these cases we have found, in the large $N$ limit, precise agreement
with the earlier literature on the DBI spike soliton \cite{
calmald,gibb,selfd,gauntlett,gauntlett2}.

As commented before, in these constructions (except in section 3.4),
there are no nontrivial supergravity fields in the ambient spacetime.
Hence these solutions are quite distinct from the configurations
arising from the dielectric effect
discussed in ref.~\cite{dielec}. The latter involves a collection
of D$p$-branes being `polarized' into a noncommutative configuration
by an external field. As well as using the nonabelian character of
the D-string theory, the essential new feature of the present
constructions is the introduction of unusual (\ie singular)
boundary conditions in the world-volume theory \cite{giveon,sethi}. 
For example, in
the BPS funnel \reef{spike} the scalars diverge at $\s=\signot$.
To comment on these boundary conditions further, let us consider
the solution for different representation of the generators $\al^i$.
Throughout the paper, we emphasized the irreducible $N\times N$
representation, for which we found that the funnel corresponded
precisely to the $N$ D-strings expanding into a single
D3-brane. One could reconsider the analysis when the $\al^i$ are
chosen as the direct sum of $q$ copies of the
${N\over q}\times{N\over q}$ representation. In this case, one
would find that the BPS funnel \reef{spike} describes
an expansion into $q$ coincident D3-branes  at $\s=\signot$.
Using energy considerations as in eq.~\reef{energy2}, naively one
might conclude that it is favorable for
this configuration to decay into the original funnel.
Of course, this is incorrect -- the $q$ D3-branes can not `decay' into
a single D3-brane. Rather one should think of the new solution
as a different superselection sector, which is distinguished in
our construction by imposing a distinct set of boundary conditions
at $\s\rightarrow\signot$ (or alternatively at $R\rightarrow\infty$).

Actually from the D-string point of view, the scalar fields start
to vary extremely rapidly as $\s\rightarrow\signot$ and so our
description in terms of the low energy world-volume action (\ref{action2})
will break down before this point is reached. On the other hand,
as $\s\rightarrow\infty$ the world-volume scalars are both slowly
varying and small, and so our formulation should give a very reliable
description of the physics. This behavior is complementary to the
D3-brane analysis. From this point of view, the world-volume
fields are slowly varying and small for large $R$, and rapidly varying
for small $R$. Thus these two approaches give complementary descriptions
for the DBI spike. 
We note that this complementarity arises because of the
`duality'\footnote{For simplicity, we set $\signot=0$ here. Further
we assume the irreducible $N\times N$ representation with large N,
so that $C\simeq N^2$. Both of these assumptions will apply throughout
the remainder of the discussion.}
\beq
R\simeq {N\ls^2\over\s}
\labell{duality}
\eeq
 between the world-volume coordinates in the two different
formulations. 

Let us try to be more precise about the ranges of validity where we
think we can trust the Born-Infeld analysis in each of these
approaches. Essentially, we must determine when we can confidently
ignore higher derivative corrections to the action arising from the
usual $\alpha'$ expansion in string theory. Schematically we would require
$\ls \prt^2 \Phi\ll\prt\Phi$. For the spike soliton on the
D3-brane, this translates into $R\gg\ls$, or using eq.~\reef{duality},
$\sigma\ll N\ls$. From the D-string funnel point
of view, we must require $\s\gg\ls$, which is equivalent to
$R\ll N\ls$. So we see that in a large $N$ limit, there is a significant
overlap region, and this explains, at least partially,
the good agreement we find between the two approaches in this regime.

Beyond the higher derivative corrections, 
the nonabelian action (\ref{finalbi},\ref{finalcs}) requires
additional higher order commutator corrections
\cite{notyet1,notyet2,notyet3}
--- see also the discussion in ref.~\cite{dielec}. Given this limitation,
we might conclude that our nonabelian D-string calculations are 
reliable only for small commutators. For the BPS funnel, we require
$\ls|\hR|\ll1$ since the commutators of the scalar fields
are characterized by the dimensionless quantity $\ls\hR$. In terms of
the physical radius, this restriction becomes $R\ll N\ls$,
which coincides with the restriction derived in the previous discussion.
A better restriction for avoiding the higher commutator
corrections is that the Taylor expansion of the square root in the action
\reef{action3} should converge rapidly. This requirement leads
to the more restrictive condition that $R\ll\sqrt{N}\ls$. However,
for large $N$, there is still overlap with the D3-brane approach
over a large region. We should add, however, that this discussion applies
to generic field corrections. There are some indications that for
supersymmetric configurations, the higher commutator corrections may vanish
\cite{notyet1,notyet2}, and so the less conservative restriction above
may be the correct one for the BPS funnel.

We should also remember that we have neglected
gravitational effects, which is justified when $gN\ll 1$. Since none
of our analysis involves the string coupling $g$, this requirement is
easily satisfied by going to very weak coupling.

In section 4, we also found remarkable agreement for the dynamics
of small fluctuations on the D-string funnel and those on
the D3-brane spike, for large $N$ and also $\ell\ll N$. Our analysis
begins to show significant discrepancies for higher $\ell$ modes.
In particular, the spectrum of modes on the noncommutative funnel
is truncated at $\ell_{\rm max}=N-1$. This brings us to the
puzzle arising from ref.~\cite{K/T}. There the detailed analysis
of the fluctuations on the D3-brane spike showed that there
was no suppression of the higher $\ell$ modes near the core.
Hence modes with arbitrarily large $\ell$ appeared to propagate
out to infinity, and the spike would seem to retain its
three-dimensional character arbitrarily far out rather making
a transition to string-like behavior. Since our D-string analysis 
provides a reliable description of physics at the core of the
spike, we conclude that this result cannot be correct.
We have found that only a finite number of modes propagate
far from the D3-brane. Note, however, that for $N$ D-strings,
this number is $N^2$, not just $N$, due to the nonabelian
character of the coincident D-strings.

Above we considered in detail in what regimes the D3-brane
and D-string descriptions would be trustworthy. However, this
analysis was only for the spike or funnel solution itself, which
plays the role of a background in the calculations of the
linearized fluctuations. Hence we should repeat this preceding analysis
for the fluctuations themselves. In particular, a fluctuation
on the D3-brane with angular momentum number $\ell$ oscillates
on spheres of constant radius with an effective wavelength
$\lambda=R/\ell$. Hence for higher derivative corrections to the
D3-brane action to be negligible, we must require that $\lambda/\ls\gg1$.
Hence $R\gg\ell\ls$ or from eq.~\reef{duality}, $\s\ll(N/\ell)\ls$.
Therefore even if we assume $N$ is large,
we can only trust the linearized equations of motion to accurately
describe the propagation of fluctuations far out on the spike for
$\ell\ll N$. Similarly the regime of validity of the analysis for fluctuations
on the D-string funnel is more restrictive for the higher $\ell$
modes. In this case, we require that higher commutator corrections
to the nonabelian Born-Infeld action remain negligible. Given
the commutators in eq.~\reef{ell}, it appears the relevant quantity
to characterize the commutators is $\ell \ls\hR$. Thus our
calculations would be trustworthy for $R\ll(N/\ell)\ls$ or
$\s\gg\ell\ls$. Hence we conclude that we should not expect the
two approaches to agree on the dynamics of the linearized
fluctuations for $\ell\sim\sqrt{N}$ or higher.

Therefore it seems that the resolution of the conflict between
the results of ref.~\cite{K/T} and the present paper is that
the dynamics of the high $\ell$ modes is significantly
altered in a transition region between regimes where either
of the two formulations can be trusted. In particular, higher
derivative corrections to the D3-brane action must play an
important role near the core of the spike, and cause the very
high $\ell$ modes to be reflected back out to the region of
large radius. Unfortunately, beyond the observations made
above, we can not
provide a detailed account of this suppression mechanism.

To summarize, we have again seen that the Born-Infeld action 
is a remarkably powerful tool in describing the low
energy dynamics of D$p$-branes. On the one hand,
with the D3-brane action one
can construct spike configurations corresponding to D-strings
attached to the D3-brane,
and the validity of these solutions seems to go far beyond naive
expectations. In this paper, we have shown how these configurations
also emerge from the D-string action in terms of noncommutative
geometry. This formulation provides a reliable description of
the central core of the DBI spike, but is also reliable to a
very large radius when the number of D-strings is large.
Hence we find surprising agreement with the original
D3-brane theory point of view. Combining these two approaches
presents an intriguing picture of D-strings attached
to an orthogonal D3-brane. At large radius we have
a continuous D3-brane being smoothly deformed into the spike
geometry. However, near the core far out along the spike, there
is a metamorphosis to a discrete structure, namely a
noncommutative funnel geometry. One can begin to gain insight into this
transition from the recent observations in ref.~\cite{noncom1}.
In the D3-brane analysis, the spike is a magnetic monopole. Thus
there is also a constant flux of magnetic field on the
spheres of constant radius surrounding the spike. As the radius
shrinks the local flux density becomes very large, and hence we can expect
to enter a regime where noncommutative geometry provides an
efficient description of the system.

Despite the striking agreement in the shape of the
D3-brane spike and the D-string funnel, one may
question whether or not the agreement should actually
be complete.
In ref.~\cite{larus}, Thorlacius showed that the
BPS spike solution found on the D3-brane was actually
a solution of the full string action. This analysis
was made for the electric spike describing a fundamental
string, and it appears the proof would be more involved for
the magnetic monopole describing the D-string \cite{laruspriv}.
Certainly the D3-brane and the D-string theories
yield the same shape for the latter configuration,
however they only agree on the overall coefficient for large $N$.
For small $N$, the coefficients will differ significantly.
It may still be that the monopole provides an exact boundary conformal
field theory, but that there is a
`renormalization' of the metric relevant to describing the geometry.
This would be somewhat similar to the difference between the closed
and open string metrics in situations where noncommutative
geometry is relevant \cite{noncom1}.
However, we will leave this question for future work.

It would be interesting to generalize the present discussion to 
other D-brane systems. The extension to D$p$-branes
ending on orthogonal D$(p+2)$-branes would follow trivially
by the application of T-duality. A more interesting extension
would be to consider a D-string ending on an orthogonal D$p$-brane
with $p\not=3$. From the lowest order equation of motion \reef{motion},
a static configuration would still have to satisfy 
\beq
\prt_\s^2\Phi^i=\left[\Phi^j,[\Phi^j,\Phi^i]\right]\ .
\eeq
With the ansatz $\Phi^i=\hR(\s)\,G^i$ for some constant
matrices $G^i$, this equation can still only yield a
differential equation of the form
\beq
\hR''= a\,\hR^3\ ,
\labell{oldnew}
\eeq
with some constant $a$, and one will again find
solutions of the form $R\propto\s^{-1}$. Hence this
type of profile is universal for all funnels on the D-string
in any situation, and the only difference
from the D3-brane funnel will be in the overall constant coefficient.
This result is slightly surprising
since from the dual D$p$-brane formulation,
one would generically expect that for large $R$, solutions will
essentially be harmonic functions
behaving like $\s\propto R^{-(p-2)}$ or $R\propto\s^{-1/(p-2)}$.
The resolution of this puzzle seems to be that 
the two profiles apply in distinct regimes, the first for
small $R$ and the second for large $R$. However, there is the
possibility that solutions of the full Born-Infeld action
will display a transition from one kind of behavior to another.
In fact, we have begun analyzing the case of a D-string ending
on an orthogonal D5-brane in detail, and we find that funnel solutions
do indeed make this kind of transition \cite{prep}.

\section*{Acknowledgments}
This research was supported by NSERC of Canada and Fonds FCAR du Qu\'ebec.
We would like to thank Curt Callan, Aki Hashimoto, Simeon Hellerman,
Clifford Johnson, Amanda Peet, 
Wati Taylor, Mark van Raamsdonk 
and L\'arus Thorlacius for useful conversations.
We are grateful to Koji Hashimoto and Simeon Hellerman for
bringing to our attention the interpretation of the nonabelian
D-string theory in terms of the Nahm equations.
NRC would also like to thank the Physics Department at UCSB for its
hospitality during the course of this work.
Research at the ITP, UCSB was supported by NSF Grant PHY94-07194.


\begin{thebibliography}{99}

\bibitem 
{Polchin}{J.~Polchinski, Phys. Rev. Lett. {\bf 75}
(1995) 4724, hep-th/9510017.}

\bibitem 
{leigh}{J. Dai, R.G. Leigh and J. Polchinski, Mod. Phys. Lett. {\bf A4} (1989)
2073.}

\bibitem 
{Polchin2}{J.~Polchinski, {\it ``TASI lectures on D-branes,''}
 hep-th/9611050;\\
J.~Polchinski, M. Chaudhuri and C.V. Johnson, {\it ``Notes on D-branes,''}
hep-th/9602052;
C.V.~Johnson,
{\it ``Etudes on D-branes,''} hep-th/9812196.}

\bibitem 
{bin}{R.G. Leigh, Mod. Phys. Lett. {\bf A4} (1989) 2767.}

\bibitem 
{mike}{M.R. Douglas, {\it ``Branes within branes,''} hep-th/9512077.}

\bibitem 
{cs}{M. Li, {Nucl. Phys.} {\bf B460} (1996) 351, hep-th/9510161;\\
M. Green, J.A. Harvey and G. Moore, Class. Quant. Grav. {\bf 14}
(1997) 47, hep-th/9605033.}

\bibitem 
{calmald}{C.G.~Callan and J.M.~Maldacena, Nucl. Phys. {\bf B513}
(1998) 109, hep-th/9708147.}

\bibitem 
{gibb}{G.W.~Gibbons, Nucl. Phys. {\bf B514} (1998) 603, hep-th/9709027.}

\bibitem 
{selfd}{ P.S.~Howe, N.D.~Lambert and P.C.~West,
Nucl.\ Phys.\ {\bf B515} (1998) 203, hep-th/9709014.}

\bibitem 
{larus}{L.~Thorlacius, Phys.\ Rev.\ Lett.\ {\bf 80} (1998) 1588, 
hep-th/9710181.}

\bibitem 
{larspeet}{S.~Lee, A.~Peet and L.~Thorlacius, 
Nucl.\ Phys.\ {\bf B514} (1998) 161, hep-th/9710097.}

\bibitem 
{fluct}{D.~Bak, J.~Lee, H.~Min, Phys.\ Rev.\ {\bf D59} (1999) 045011,
hep-th/9806149;\\
K.~Savvidy, G.~Savvidy, {\it ``Neumann boundary conditions from
Born-Infeld dynamics,''} hep-th/9902023.}

\bibitem 
{K/T}{D.~Kastor and J.~Traschen, {\it ``Dynamics of the DBI spike
  soliton,''} hep-th/9906237.}

\bibitem 
{ded}{D.-E.~Diaconescu, Nucl.\ Phys.\ {\bf B503} (1997) 220,
hep-th/9608163.}

\bibitem 
{gauntlett}{J.P.~Gauntlett, J.~Gomis, P.K.~Townsend, 
{\it ``BPS bounds for worldvolume branes,''} hep-th/9711205;\\
D.~Brecher, Phys.\ Lett.\ {\bf B442} (1998) 117, hep-th/9804180.}

\bibitem 
{giveon}{A.~Giveon and D.~Kutasov, Rev.\ Mod.\ Phys.\ {\bf 71} (1999)
983, hep-th/9802067.}

\bibitem 
{sethi}{A.~Kapustin and S.~Sethi, Adv.\ Theor.\ Math.\ Phys.\  {\bf 2}
(1998) 571, hep-th/9804027;\\
D.~Tsimpis, Phys.\ Lett.\ {\bf B433} (1998) 287, hep-th/9804081.}

\bibitem 
{hashimoto}{K.~Hashimoto, Prog.\ Theor.\ Phys.\ {\bf 101} (1999) 1353,
hep-th/9808185.}

\bibitem 
{gorsky}{A.~Gorsky and K.~Selivanov, {\it ``Junctions and the fate of
branes in external fields,''} hep-th/9904041.}

\bibitem 
{nahm}{W.~Nahm, Phys.\ Lett.\ {\bf B90} (1980) 413;\\
W.~Nahm, {\it ``The construction of all self-dual multimonopoles by the
ADHM method,''} in ``Monopoles in quantum field theory,'' Craigie et
al.\ (eds), World Scientific, Singapore (1982).}

\bibitem 
{dielec}{R.C.~Myers, {\it ``Dielectric-branes,''} hep-th/9910053}

\bibitem 
{watiprep}{W.~Taylor and M.~Van Raamsdonk,
{\it ``Multiple D$p$-branes in weak background fields,''}
hep-th/9910052.}

\bibitem 
{roberto}{R. Emparan, Phys. Lett. {\bf B423} (1998) 71, hep-th/9711106.}

\bibitem 
{aki}{A.~Hashimoto, Phys.Rev. {\bf D57 (1998)} 6441, hep-th/9711097.}

\bibitem 
{gauntlett2}{J.~Gauntlett, C.~Koehl, D.~Mateos, P.~Townsend,
M.~Zamaklar, Phys.\ Rev.\ {\bf D60} (1999) 045004, hep-th/9903156.}

\bibitem 
{ghoroku}{K.~Ghoroku, K.~Kaneko, {\it ``Born-Infeld strings between
D-branes,''} hep-th/9908154.}

\bibitem 
{khashimoto}{K.~Hashimoto, {\it ``Born-Infeld dynamics in 
uniform electric field,''} hep-th/9905162.}

\bibitem 
{yet2}{A.A.~Tseytlin,
{\it ``Born-Infeld action, supersymmetry and string theory,''}
hep-th/9908105.}

\bibitem 
{wati2}{D.~Kabat and W.I.~Taylor,
Adv. Theor. Math. Phys. {\bf 2} (1998) 181, hep-th/9711078;\\
S.~Rey, {\it ``Gravitating M(atrix) Q balls,''}
hep-th/9711081.}

\bibitem 
{rocek}{M.~Rocek and A.A.~Tseytlin,
Phys.\ Rev.\ {\bf D59} (1999) 106001, hep-th/9811232.}

\bibitem 
{watirev}{W.~Taylor,
{\it ``Lectures on D-branes, gauge theory and M(atrices),''}
at 2nd Trieste Conference on Duality in String
Theory, Trieste, Italy, 16-20 Jun 1997,
hep-th/9801182.}

\bibitem 
{yet}{A.A. Tseytlin, {Nucl. Phys.} {\bf B501} (1997) 41,
hep-th/9701125.}

\bibitem 
{prasad}{S.A.~Brown, H.~Panagopoulos and M.K.~Prasad, Phys.\ Rev.\
{\bf D26} (1982) 854;\\
H.~Panagopoulos, Phys.\ Rev.\ {\bf D28} (1983) 380.}

\bibitem 
{son}{N.R.~Constable, S.~Hellerman and \O.~Tafjord, work in progress.}

\bibitem 
{pqsting}{J.H.~Schwarz,
Phys.\ Lett.\ {\bf B360} (1995) 13, hep-th/9508143;
Erratum-ibid. {\bf B364} (1995) 252;\\ 
E.~Witten, Nucl.\ Phys.\ {\bf B460} (1996) 335, hep-th/9510135;\\
M.~Li,
Nucl.\ Phys.\ {\bf B460} (1996) 351, hep-th/9510161.}

\bibitem 
{solute}{M.J.~Duff, R.R.~Khuri and J.X.~Lu,
Phys.\ Rept.\ {\bf 259} (1995) 213, hep-th/9412184;\\
G.T.~Horowitz and A.~Strominger,
Nucl.\ Phys.\ {\bf B360} (1991) 197.}

\bibitem 
{fuzzball}{J.~Madore,
Class.\ Quant.\ Grav.\ {\bf 9} (1992) 69;
Ann.\ Phys.\ {\bf 219} (1992) 187;
Phys.\ Lett.\ {\bf B263} (1991) 245;\\
J. Hoppe, MIT Ph.D.\ thesis, 1982;
Elem.\ Part.\ Res.\ J.\ (Kyoto) {\bf 80} (1989) 145.}

\bibitem 
{notyet1}{A. Hashimoto and W. Taylor, Nucl. Phys. {\bf B503} (1997) 193,
hep-th/9703217.}

\bibitem 
{notyet2}{P.~Bain, {\it ``On the non-abelian Born-Infeld action,''}
hep-th/9909154.}

\bibitem 
{notyet3}{C.P. Bachas and P. Bain, unpublished; A. Hashimoto and
R.C. Myers, unpublished.}

\bibitem 
{noncom1}{N.~Seiberg and E.~Witten,
JHEP {\bf 09} (1999) 032, hep-th/9908142.}

\bibitem 
{laruspriv}{L.~Thorlacius, private communication.}

\bibitem 
{prep}{N. Constable, \O. Tafjord and R.C. Myers, in preparation.}





\end{thebibliography}
\end{document}